\documentclass[aps,prd,secnumarabic,amssymb, amsmath,nobibnotes,nofootinbib,11pt]{revtex4} 
\usepackage{amsfonts,amsmath,hyperref,url, color}
\usepackage{bm, bbm}
\usepackage{graphicx}
\usepackage{mathtools}
\usepackage{bbold}

\newcommand{\be}{\begin{equation}}
\newcommand{\bal}{\begin{align}}
\newcommand{\eal}{\end{align}}
\newcommand{\ee}{\end{equation}}
\newcommand{\bea}{\begin{eqnarray}}
\newcommand{\eea}{\end{eqnarray}}
\newcommand{\bit}{\begin{itemize}}
\newcommand{\eit}{\end{itemize}}

\newcommand{\ba}{\begin{aligned}}
\newcommand{\ea}{\end{aligned}}

\usepackage{color}

\usepackage[utf8]{inputenc}
\begin{document}

\title{Quantum particles in non-commutative space-time: an identity crisis}
\author{Michele Arzano}
\email{michele.arzano@na.infn.it}
\affiliation{Dipartimento di Fisica ``E. Pancini", Universit\`a di Napoli Federico II, I-80125 Napoli, Italy\\}
\affiliation{INFN, Sezione di Napoli,\\ Complesso Universitario di Monte S. Angelo,\\
Via Cintia Edificio 6, 80126 Napoli, Italy}
\author{Jerzy Kowalski-Glikman}
\email{jerzy.kowalski-glikman@uwr.edu.pl}
\affiliation{University of Wroc\l{}aw, Faculty of Physics and Astronomy \\ Maksa Borna 9, 50-204 Wroc\l{}aw, Poland\\}
\affiliation{National Centre for Nuclear Reserach,\\Pasteura 7, 02-093 Warsaw, Poland}


\begin{abstract}
We argue that the notion of identical particles is no longer well defined in quantum systems governed by non-commutative deformations of space-time symmetries. Such models are characterized by four-momentum space given by a non-abelian Lie group. Our analysis is based on the observation that, for states containing more than one particle, only the total momentum of the system is a well defined quantum number. Such total momentum is obtained from the non-abelian composition of the particles individual momenta which are no longer uniquely defined. The main upshot of our analysis is that all previous attempts to construct Fock spaces for these models rested on wrong assumptions and indeed have been unsuccessful. We also show how the natural {\it braiding} of momentum quantum numbers which characterizes the exchange of factors in the tensor product of states is covariant under relativistic transformations thus solving a long standing problem in the field. 
\end{abstract}

\maketitle

\newpage

Indistinguishability of identical particles is, like entanglement, a characterizing feature of quantum systems that has no counterpart in the classical realm and, in fact, is at the basis of the radical difference between the statistical mechanics of quantum systems compared to that of classical systems. 

In relativistic quantum theories the behaviour of multi-particle systems is dictated by the spin-statistics theorem, a pillar of local quantum field theory, which states that bosons and fermions are described by a Fock space contructed from symmetrized and antisymmetrized tensor products of one-particle states respectively.

Since unitary representations of space-time isometries are at the basis of the spin-statistics connection (see e.g. \cite{Marletto:2021lhv} and references therein), it would be natural to contemplate possible departures from such paradigm 
when quantum features of space-time enter the game. Speculations concerning violations of the spin-statistics theorem in quantum gravity have appeared in the literature (see e.g. \cite{Jackson:2007tn}, \cite{Swain:2008fi}) lacking however a specific field theoretic model which could explicitly realize the envisioned non-standard behaviour of multi-particle states. 

As it turns out departures from the ordinary description of multi-particle states in terms of elements of bosonic or fermionic Fock spaces appear to be unavoidable in certain non-commutative deformations of quantum field theories in which the usual four-momentum space is replaced by a non-abelian Lie group. The particular model we have in mind here is a quantum field theory based on representations of the $\kappa$--Poincar\'e algebra Hopf algebra \cite{Lukierski:1991pn,Lukierski:1992dt,Kosinski:1999ix} which, in the configuration space formulation, is built on fields on the non-commutative $\kappa$-Minkowski space-time \cite{Lukierski:1993wx,Majid:1994cy,Dimitrijevic:2003wv,Bu:2006dm,Arzano:2007gr,Meljanac:2007xb,Freidel:2007hk,Arzano:2007nx,Arzano:2009ci,Meljanac:2010ps,Arzano:2020jro} and in the momentum space formulation is described by fields on a non-abelian Lie group which supports a transitive action of the Lorentz group \cite{Amelino-Camelia:2001rtw,Arzano:2010jw,Arzano:2017uuh,Poulain:2018two,Arzano:2018gii} (see \cite{Arzano:2021hpg} and \cite{Arzano:2021scz} for recent comprehensive reviews on the subject).

It was pointed out some time ago \cite{Arzano:2007ef} that for such theories, as we will briefly review below, a formulation of multi-particle states in terms of an ordinary Fock space construction turns out to be problematic. To date, despite vartious attempts in the literature (see e.g. \cite{Daszkiewicz:2007ru,Young:2007ag,Govindarajan:2008qa,Young:2008zg,Young:2008zm,Arzano:2008bt,Kim:2009jk,Govindarajan:2009wt}), no consistent, fully satisfactory formulation for multi-particle sates in $\kappa$-deformed quantum field theory has been laid out. In this letter we propose a new framework for describing systems of many particles in such theories taking inspiration from the formal analogy between the group-valued momenta carried by the deformed one-particle states and the quantum numbers characterizing non-abelian anyons \cite{Lo:1993hp}. Our treatment is based on the introduction a non-trivial {\it braiding} when exchanging the factors of the tensor product of two one-particle states, an operation which is covariant under the deformed action of the Lorentz group on group-valued momenta. Such braiding is formally analogous to the phenomenon of \textit{flux-metamorphosis} for non-abelian anyons (see e.g. \cite{Lo:1993hp}). The main upshot of our analysis is that the efforts to find a consistent construction of (anti)-symmetrized states in $\kappa$-deformed quantum field theory have been unsuccessful so far since the very notion of {\it identical particle} is no longer well defined when momentum quantum numbers are elements of a non-abelian group.\\

In ordinary Minkowski spacetime quantum field theory {\it one-particle states} are elements of a Hilbert space $\mathcal{H}$ which carries an irreducible representation of the Poincar\'e group \cite{Weinberg:1995mt}. For a scalar field such space can be described by functions of four-momenta restricted to the positive mass-shell $p^2-m^2=0$. In usual textbook treatments a basis for such space is given in terms of eigenstates of the translation generators denoted by kets $| p\rangle$ 
\be
P^{\mu}\, | p\rangle = p^{\mu}\, | p\rangle
\ee
with $p^0=\sqrt{\vec{p}^2+m^2}$. Indistinguishability of the particle excitations of the theory requires that, for example, the state of two (bosonic) particles should be represented by the {\it symmetrized tensor product}
\be
|p_1, p_2\rangle \equiv \frac{1}{\sqrt{2}}\left(| p_1\rangle \otimes | p_2\rangle + | p_2\rangle \otimes | p_1\rangle\right)\,.
\ee
Generalizing to $n$-particles their Hilbert space will be given by the symmetrized space
\begin{equation}
S_n\mathcal{H}^n=\frac{1}{\sqrt{n!}}\sum_{\sigma\in P_n}\sigma(\mathcal{H}^{\otimes n})\, .
\end{equation}
where $\sigma$ is an element of the permutation group of $n$-elements $P_n$ and $\mathcal{H}^{\otimes n}$ is the $n$-fold tensor product of $n$-copies of $\mathcal{H}$. The Hilbert space comprising all possible multiparticle configurations of the theory is given by a direct sum of these $n$-particle Hilbert spaces i.e. the Fock space
\begin{equation}
\mathcal{F}_{s}(\mathcal{H})=\bigoplus_{n=0}^{\infty} S_n\mathcal{H}^n\,,
\end{equation}
where $\mathcal{H}^0=\mathbb{C}$ and the subscript ‘‘s" stands for ‘‘symmetrized", for a fermionic field one applies the same procedure using {\it anti-symmetrized} tensor products of one-particle states.

As it turns out such bosonic and fermionic Fock spaces are not the only representations of the Poincaré group which can be used to describe multiparticle systems. It is indeed well known that in $2+1$ space-time dimensions there exist {\it anyonic} particle excitations \cite{Leinaas:1977fm,Wilczek:1982wy} which are not bosonic nor fermionic and for which one has to resort to alternative Hilbert space formulations for the space of states \cite{Mund:1993cf}. Much less known, at least until recently \cite{Csaki:2020yei}, is that also in ordinary quantum field theory in $3+1$-dimensional Minkowski space-time more general multiparticle representations of the Poincaré group are possible. These occur when considering asymptotic states for the scattering of particles carrying electric and magnetic charges (dyons). As noted by Zwanziger many years ago \cite{Zwanziger:1972sx} aymptotic states for such particles carry an additional angular momentum quantum number which can be captured by multi-particle states carrying a pair-wise helicity which are not simple tensor products of their one-particle components \cite{Csaki:2020yei}. Thus asymptotic states for such dyons can not be described in terms of elements of a Fock space.

The non-commutative models we discuss in this letter provide yet another example of quantum field theories for which a standard formulation of multiparticle states in terms of a Fock space fails. We focus here on the quantum states of particles described by a field theory constructed from representations of the $\kappa$--Poincar\'e algebra Hopf algebra. The main feature of the model is that particles four-momenta are described by elements of a Lie group obtained by exponentiating the Lie algebra 
\be\label{kMink}
[X^0, X^i] = \frac{i}{\kappa} X^i\,, \quad [X^i, X^j] = 0\,; \quad i,j= 1,\ldots,3
\ee
known in the non-commutative field theory literature as the $\kappa$-Minkowski space-time and in the mathematical literature as the $\mathfrak{an}(3)$ Lie algebra. The parameter $\kappa$, with dimensions of inverse length, is usually taken as a UV scale identified with the Planck energy.  The Lie group obtained by exponentiating the algebra above is denoted by $AN(3)$ and it can be seen as the momentum space labelling non-commutative plane waves on the $\kappa$-Minkowski space-time. This curved momentum space, the group manifold of the $AN(3)$ group,   has the geometry of (a half of) de Sitter space (see \cite{Arzano:2021hpg} and \cite{Arzano:2021scz} for details). The plane wave, or an element $g\in AN(3)$ can be parametrized as
\be\label{Kexp}
g = e^{i p_i X^i} e^{i p_0 X^0}\,.
\ee 
In this parametrization $p_0$, $p_i$ are known as {\it bicrossproduct coordinates} on $AN(3)$ \cite{Majid:1994cy}. The matrix representation of this group is discussed in details in the recent paper \cite{Arzano:2022ewc}. 

One of the characterizing features of the kinematics associated to such model is the non-abelian composition of four-momenta which follows directly from the non-abelian nature of the $ AN(3)$ group. Using the commutation relations of the $\mathfrak{an}(3)$ Lie algebra or the explicit matrix form of the group elements one can write
\begin{equation}
        e^{ip_{i}X^{i}}e^{ip_{0}X^{0}}=e^{ip_{0}X^{0}}e^{ip_{i}e^{p_{0}/\kappa}X^{i}}
\end{equation}
from which it follows that
\begin{multline}
e^{ip^{(1)}X} e^{ip^{(2)}X} =    e^{ip_{i}^{(1)}X^{i}}e^{ip^{(1)}_{0}X^{0}}e^{ip_{i}^{(2)}X^{i}}e^{ip^{(2)}_{0}X^{0}}=e^{ i(p_{i}^{(1)}+p_{i}^{(2)}e^{-p_{0}^{(1)}/\kappa})X^{i}}e^{i(p_{0}^{(1)}+p_{0}^{(2)})X^0}\equiv e^{ i( p^{(1)} \oplus p^{(2)}) X}
\end{multline}
which determines the non-abelian addition
\be\label{oplus}
p^{(1)} \oplus p^{(2)} = \left(p_{0}^{(1)}+p_{0}^{(2)},\mathbf{p}^{\,(1)}+\mathbf{p}^{\,(2)}e^{-p_{0}^{(1)}/\kappa}\right)\,.
\ee
The inverse plane wave
\begin{align}
 (e^{ipX})^{-1} =  \left(e^{ip_{i}X^{i}}e^{ip_{0}X^{0}}\right)^{-1}=e^{-ip_{i}e^{p_{0}/\kappa}X^i}e^{-ip_{0}X^{0}} \equiv  e^{i (\ominus p) X}
\end{align}
determines the operation 
\be
\ominus p \equiv \left(-p_{0}, -\mathbf{p}\, e^{p_{0}/\kappa}\right)
\ee
which by construction ensures that $p\oplus (\ominus(p)) = 0$.

For our purposes it is important to stress that the deformed addition rule \eqref{oplus} reflects the non-trivial Hopf algebra structure of $\kappa$-Poincaré,  namely the fact that its elements act on tensor product representations via a generalization of the Leibniz rule. For the tensor product of two representations the non-abelian addition of momenta can be recast in terms of the non-trivial {\it coproduct} for the generators of time and space translation generators
\begin{align}\label{cprod}
 \triangle(P_{i})=P_i\otimes \mathbbm{1} +e^{-\frac{P_0}{\kappa}}\otimes P_i\,,\quad
  \triangle (P_{0}) = P_{0}\otimes \mathbbm{1} +  \mathbbm{1} \otimes P_{0}.
\end{align}


This observation brings us directly to the core of our problem. As in ordinary quantum field theory one-particle states in the $\kappa$-deformed context will carry a representation of the $\kappa$-Poincaré algebra. We will denote such states with kets $|g\rangle$ labelled by elements of the group $g\in AN(3)$. The action of the translations generators $P_{\mu}$ on such states is 
\be
P_{\mu}\, |g\rangle = p_{\mu}(g)\, |g\rangle
\ee
where $p_{\mu}(g)$ are the bicrossproduct coordinates assciated to the group element $g$ (see  \cite{Arzano:2022ewc} for a discussion of the action of the other elements of the $\kappa$-Poincaré algebra on such states). Let us consider a two-particle state
\be
|g\rangle \otimes |h\rangle\,.
\ee
The total four-momentum carried by such state is given by the product of two group elements $gh$ and, once a parametrization of the group is chosen, it leads to a non abelian composition rule like \eqref{oplus}. Such total four-momentum is also the eigenvalue of the translations generators acting on the tensor product via a deformed Lebniz rule \eqref{cprod}, indeed for the bicrospproduct generators $P_{\mu}$ we have that
\be
\triangle(P_{\mu}) (|g\rangle \otimes |h\rangle) =p_{\mu}(gh) (|g\rangle \otimes |h\rangle)\,.
\ee
with $p_{\mu}(gh)$ the coordinates of the group element $gh$.

As recalled above when working with indistinguishable particles in ordinary quantum field theory the tensor product of two one-particle states has to be symmetrized or anti-symmetrized if one is dealing with bosons or fermions respectively. Applying a naive symmetrization to the state above would lead to the two-particle state
\be\label{nsy}
\frac{1}{\sqrt{2}}\left(|g\rangle \otimes |h\rangle + |h\rangle \otimes |g\rangle\right)
\ee
which is not an eigenstate of translation generators since the two states $|g\rangle \otimes |h\rangle$ and $|h\rangle \otimes |g\rangle$ have different total momenta $p_\mu(gh)$ and $p_\mu(hg)$ respectively. In other words such state does not have a well defined total momentum. One could opt to give up the requirement of having a well defined total momentum (as done for example in \cite{Noui:2006kv}) but the naive symmetrization \eqref{nsy} is problematic per se since it contradicts the very assumption of indistinguishability. Indeed {\it if two particles are indistinguishable swapping the factors in the tensor product describing their state should lead to another state which is indistinguishable from the original one}, i.e. with the same quantum numbers. Clearly this is not the case when working with group-valued momenta since the momentum carried by the two-particle state $|g\rangle \otimes |h\rangle$
is different from the momentum carried by the state $|h\rangle \otimes |g\rangle$.\\

We now make a key observation. Let us notice that if a swapping of the factors in the tensor product of two one-particle states is accompanied by a change in the four-momentum 
\be\label{2p2}
|g\rangle \otimes |h\rangle\, \longrightarrow\, |g h g^{-1}\rangle \otimes |g\rangle\,.
\ee
such state has the same momentum as the original state $|g\rangle \otimes |h\rangle$. This 
phenomenon is not new and indeed in lower dimensional physics it occurs for non-abelian anyons and it is called ‘‘flux metamorphosis" see e.g. \cite{Lo:1993hp}. Now, requiring that the two particles be identical would naturally lead to consider the symmetrized state
\be\label{symms}
 |g\rangle \otimes |h\rangle +|g h g^{-1}\rangle \otimes |g\rangle \,,
\ee
with a well-defined total momentum $gh$. However it is evident that we could swap particles in the state \eqref{2p2} once more to obtain a {\it different} state 
\be
|(gh)g (gh)^{-1}\rangle \otimes |g h g^{-1}\rangle\,,
\ee
with the same total momentum $gh$ and so on if one keeps exchanging the particles. This renders the choice of symmetrized state \eqref{symms} completely arbitrary since any state obtained from any number of swappings of the tensor product factors could be superimposed to $|g\rangle \otimes |h\rangle$ leading to a state with the same total momentum. We conclude that all possible tensor product states obtained from the ‘‘braided" swapping of factors in the tensor product of one-particle states are legitimate representatives a two-particle state. The analogy with non-abelian anyons suggests that only the total momentum of the system is a well defined quantum number and the notion of identical particles is not well defined. Indeed at the mathematical level the four-momentum of such particles is an element of a non-abelian Lie group like the flux quantum number carried by non-abelian anyons. These considerations suggest that a Fock space description for quantum particles in $\kappa$-Poincaré might not be possible.\\ 

Let us stress that the formal analogy between particles with group-valued momenta and non-abelian anyons we invoked for our analysis is not completely new. Indeed in $2+1$ space-time dimensions it has been known since the 1990s (see e.g. \cite{Matschull:1997du} and references therein) that the momentum space of particles coupled to gravity (which in this case is a topological theory and it does not admit propagating degrees of freedom) is given by the non-abelian group $SL(2,\mathbb{R})$. The fact that such Lie group momentum space is associated to a Hopf algebra deformation of the three-dimensional Poincaré group was noticed in \cite{Bais:1998yn} and explored in detail in \cite{Bais:2002ye}. These works also pointed out the anyon-like features appearing in the study of the scattering of these topologically gravitating particles and the possibility of a braiding of the momentum quantum numbers like in \eqref{2p2} when two particles are exchanged. In \cite{Arzano:2013sta} we proposed a description of multi-particle states for a quantum field theory of particles with $SL(2,\mathbb{R})$ momentum space and associated deformed symmetries making use of such braiding.\\

A crucial issue one has to address when considering the non-trivial braiding we just introduced is that of its Lorentz covariance. 
In the case of the $\kappa$-Poincaré algebra the action of the Lorentz group on the $AN(3)$ momentum space is more subtle than the $2+1$-dimensional case mentioned above \cite{Arzano:2022ewc}. Such action is, in fact, determined by the Iwasawa decomposition of the Lie group $SO(4,1)$ as follows. Let us focus on the Iwasawa decomposition $SO(4,1)= SO(3,1) AN(3) $. A given element $G$ of the group $SO(4,1)$ can be decomposed as
\be
SO(4,1)\ni G=\Lambda\, g \in SO(3,1) AN(3) 
\ee
where $\Lambda\in SO(3,1)$ and $g\in AN(3)$. We can also wright the ‘‘right" Iwasawa decomposition of the same element
\be
G= \Lambda\, g = g' \Lambda'_{g} \in  AN(3) SO(3,1) 
\ee
and thus define the Lorentz transformed $AN(3)$ momentum as
\be\label{lorentza1}
g' = \Lambda\, g\, \Lambda^{'-1}_{g}\,, 
\ee
so that given Lorentz transformation $\Lambda$ and the $AN(3)$ group element $g$ we can uniquely construct its Lorentz transformation $g'$ and the compensating transformation $\Lambda'_g$. 
When written in infinitesimal form this action reproduces the commutator between the generators of the Lorentz group and the ones of space-time translations for any given choice of coordinates on $AN(3)$ i.e. for any {\it basis} of the $\kappa$-Poincaré algebra (see \cite{Arzano:2022ewc} for details). 

Consider now an $AN(3)$ group element being a product $gh$. According to the construction presented above, its Lorentz transformation under the action of the Lorentz group element is defined to be
\begin{align}
    (gh)' = \Lambda gh \Lambda_{gh}^{'-1}
\end{align}
It turns out that, contrary to the 2+1 dimensional case $(gh)'\neq g'h'$; instead
\begin{align}\label{ghprime}
    (gh)' =  \Lambda g \Lambda_g^{'-1} \, \Lambda'_g\, h \Lambda_{gh}^{'-1} = g' \, \Lambda'_g\, h \Lambda_{gh}^{'-1} 
\end{align}
It can be further shown \cite{Arzano:2022ewc} that when the Lorentz transformation is a pure rotation, $\Lambda = R$ the compensating transformation is $R'_g=R$, reflecting the triviality of the rotation coproduct in $\kappa$-Poincar\'e algebra. In this case \eqref{ghprime} reduces to the adjoint action
\begin{align}
    (gh)' = R g R^{-1}\, R h R^{-1} = g'  h' 
\end{align}

As shown in \cite{Arzano:2022ewc} the  coproduct  of Lorentz transformation on a two-particle state can be derived from the relation \eqref{ghprime}. It follows that the action of Lorentz transformation $\Lambda$ on a two-particle state takes the form 
\begin{align}\label{two-pL}
 |g\rangle \otimes |h\rangle \longrightarrow \Lambda \vartriangleright  |g\rangle \otimes |h\rangle=|g' \rangle\otimes|\Lambda_g \,h\,  \Lambda_{gh}^{'-1}\rangle   
\end{align}
and \eqref{ghprime} secures the covariance under Lorentz transformations, 
\begin{align}\label{DeltaPcov}
    \Delta P_\mu\left(\Lambda \vartriangleright  |g\rangle \otimes |h\rangle\right)= p_\mu((gh)')\left(\Lambda \vartriangleright  |g\rangle \otimes |h\rangle\right)
\end{align}
meaning that the total momentum of the Lorentz transformed state is the Lorentz transformation of total momentum. 

Since the group elements $g$ and $h$ in \eqref{two-pL} are arbitrary, it can be immediately generalized to the swapped state \eqref{2p2}
\begin{align}\label{two-pswL}
 |ghg^{-1}\rangle \otimes |g\rangle \longrightarrow \Lambda \vartriangleright  |ghg^{-1}\rangle \otimes |g\rangle=|(ghg^{-1})'\, \Lambda_{ghg^{-1}}^{'-1}\, \rangle\otimes |\Lambda^{-1}\,g'\, \Lambda'_g \Lambda_{gh}^{'-1}\rangle   
\end{align}
and from \eqref{DeltaPcov} we have
\begin{align}\label{DeltaPcov1}
    \Delta P_\mu\left(\Lambda \vartriangleright  |ghg^{-1}\rangle \otimes |g\rangle\right)= p_\mu((gh)')\left(\Lambda \vartriangleright  |ghg^{-1}\rangle \otimes |g\rangle\right)
\end{align}
 verifying the consistency of our framework and the covariance of the braiding.

It is useful at this point to comment on the issue of covariance of the non-trivial braiding of momentum quantum numbers under particle exchange. Indeed the unsuccessful search (see \cite{Arzano:2007ef,Daszkiewicz:2007ru,Young:2007ag,Govindarajan:2008qa,Young:2008zg,Young:2008zm,Arzano:2008bt,Kim:2009jk,Govindarajan:2009wt}) for such a braiding was one of the main obstacles encountered in the attempts of formulating a consistent Fock space for $\kappa$-deformed fields. So what makes the braiding introduced in this note so different from the ones previously explored in the literature allowing to circumvent such covariance problem? To answer this question let us write the labels in the state $|g h g^{-1}\rangle \otimes |g\rangle$ explicitly in terms of bicrossproduct coordinates $(p_0; \mathbf{p})$ on $AN(3)$ \eqref{Kexp}. In particular if $(\mathbf{p},p_0)$ is the  momentum associated to $|g\rangle$  and $(q_0;\mathbf{q})$ is the  momentum associated to $|h\rangle$ the state $|g h g^{-1}\rangle$ will carry spatial momentum $\mathbf{p} \oplus \mathbf{q} \oplus (\ominus \mathbf{p})$ and thus
\be\label{brbicr}
|g h g^{-1}\rangle \otimes |g\rangle \equiv |(q_0; \mathbf{p}\, (1-e^{-q_0/\kappa}) + e^{-p_0/\kappa}\mathbf{q})\,\rangle \otimes |(p_0;\mathbf{p})\rangle\,,
\ee
so the braiding involves {\it non-abelian sum and difference of the momenta}. The issues with the other types of braiding appeared in the literature essentially boil down to the fact that all these works were looking for a braiding which could be expressed in terms of the action of a deformed flip operator constructed from the elements of (extensions of) the $\kappa$-Poincaré algebra rather than working with the action of the momentum space group on itself. 

To be more specific let us look for example at the braiding proposed in \cite{Arzano:2008bt} given by
\be\label{brab}
|\mathbf{p}\rangle \otimes |\mathbf{q}\rangle\, \longrightarrow |e^{-p_0/\kappa}\, \mathbf{q}\,\rangle \otimes |e^{q_0/\kappa} \mathbf{p}\rangle\,.
\ee
This braiding was chosen because it provides a representation of the symmetric group on the  space of tensor products of one particle states and because it can be written in terms of an operator
\be \label{twist-p}
\mathcal{F} = e^{\frac{1}{\kappa}P_0\otimes P_j \frac{\partial}{\partial P_j}}\ . 
\ee 
reminiscent of the twist operator 
governing the deformation of the Poincaré algebra \cite{Chaichian:2004za} associated to the canonical non-commutative space-time \cite{Doplicher:1994zv} $[x_{\mu},x_{\nu}]=i\theta_{\mu\nu}$.
Such deformation known as $\theta$-Poincaré is a so-called quasi-triangular Hopf algebra and the twist operator can be used to construct a braided representation of the symmetric group. The quasi-triangular structure ensures the full compatibility between the "twisted statistics" and the action of $\theta$-Poincaré symmetry generators (see \cite{akofor} for more details for quantum fields covariant under the $\theta$-Poincaré algebra and their deformed statistics). Returning to the $\kappa$-deformed context it is easy to see that, besides the many desirable properties, the braiding \eqref{brab} is not covariant under the action of the $\kappa$-Poinacré algebra. This problem can be traced back to the fact the twist-like oparator \eqref{twist-p} does not commute with the co-product of the generator of boosts as discussed in detail in \cite{Arzano:2008bt}. Alternative approaches requiring covariance of the braiding from the outset have also been attempted. Most notably \cite{Young:2008zg} one can try to construct a covariant braiding making  use of the classical $r$-matrix of $\kappa$-Poincar\'e (see \cite{r-matrix})
$r\equiv i(N_j\otimes P_j - P_j\otimes N_j)$ 
where $N_j$ are generators of boosts, to define a twist map. It turns out that such construction can be carried out smoothly at leading order in the inverse deformation parameter $1/\kappa$. But that's the most one can do. Indeed as proved in \cite{Young:2008zm} a twist operator constructed from the $\kappa$-Poincar\'e classical $r$-matrix valid at all orders in $1/\kappa$ would not be compatible with the $\kappa$-Poincaré algebra. 

As our considerations above show these approaches failed in one way or another because the very notion of symmetrized multiparticle states loses meaning when momentum quantum numbers are element of a non-abelian Lie group. 

It is interesting to notice the similarity between the description of $\kappa$-deformed multi-particle states and that for particles carrying both electric and magnetic charges as discussed in \cite{Csaki:2020yei}. In the latter case the need to go beyond the description of states in terms of symmetrized tensor products can be traced back to the presence of an additional quantum number for systems with more than one particle: the {\it pairwise helicity} \cite{Csaki:2020yei}. In our case the additional information carried by multiparticle states is encoded in the order in which momenta are combined \cite{Arzano:2008yc}. Such analogy between quite different settings (in one the non-trivial structures are due to the presence of a long-range interaction i.e. an IR effect, in the other to a UV deformation of the kinematics) is quite fascinating and surely worth of being further understood. We leave this task to future studies.

\section*{Acknowledgment} 

For JKG, this work was supported by funds provided by the National Science Center, project number  2019/33/B/ST2/00050. JKG thanks for hospitality Perimeter Institute where this project was completed. Research at Perimeter Institute for Theoretical Physics is supported in part by the Government of Canada through NSERC and by the Province of Ontario through MRI.

This work contributes to the European Union COST Action CA18108 {\it Quantum gravity phenomenology in the multi-messenger approach.}

\end{document}